\pdfoutput=1

\documentclass[12pt,a4paper]{article}

\usepackage{ifthen} 
\newboolean{pdflatex}
\setboolean{pdflatex}{true} 

\newboolean{articletitles}
\setboolean{articletitles}{true} 

\newboolean{uprightparticles}
\setboolean{uprightparticles}{false} 

\newboolean{inbibliography}
\setboolean{inbibliography}{false} 

\usepackage{microtype}
\usepackage{lineno}  
\usepackage{xspace} 
\usepackage{caption} 

\usepackage{graphicx}  
\usepackage{color}
\usepackage{colortbl}
\graphicspath{{./figs/}} 

\usepackage{amsmath} 
\usepackage{amssymb}
\usepackage{amsfonts}
\usepackage{upgreek} 

\usepackage{bigstrut}
\usepackage{multirow}

\newcommand*\patchAmsMathEnvironmentForLineno[1]{%
\expandafter\let\csname old#1\expandafter\endcsname\csname #1\endcsname
\expandafter\let\csname oldend#1\expandafter\endcsname\csname
end#1\endcsname
 \renewenvironment{#1}%
   {\linenomath\csname old#1\endcsname}%
   {\csname oldend#1\endcsname\endlinenomath}%
}
\newcommand*\patchBothAmsMathEnvironmentsForLineno[1]{%
  \patchAmsMathEnvironmentForLineno{#1}%
  \patchAmsMathEnvironmentForLineno{#1*}%
}
\AtBeginDocument{%
\patchBothAmsMathEnvironmentsForLineno{equation}%
\patchBothAmsMathEnvironmentsForLineno{align}%
\patchBothAmsMathEnvironmentsForLineno{flalign}%
\patchBothAmsMathEnvironmentsForLineno{alignat}%
\patchBothAmsMathEnvironmentsForLineno{gather}%
\patchBothAmsMathEnvironmentsForLineno{multline}%
\patchBothAmsMathEnvironmentsForLineno{eqnarray}%
}

\usepackage{hyperref}    
\usepackage[all]{hypcap} 

\usepackage{xspace} 
\usepackage{upgreek}

\def\deta {\Delta\eta}
\def\dphi {\Delta\phi}
\def\pp{\ensuremath{pp}\xspace}

\def\pa{\ensuremath{p\text{Pb}}\xspace} 

\def\hijing {\mbox{\textsc{Hijing}}\xspace}

\def\ppb{\mbox{\ensuremath{p\!+\!\mathrm{Pb}}}\xspace}
\def\pbp{\mbox{\ensuremath{\mathrm{Pb}\!+\!p}}\xspace}

\def\sqrtnn{\ensuremath{\sqrt{s_{\scriptscriptstyle N\!N}}}}
\def\nhitvelo{\ensuremath{\mathcal N^{\mathrm{hit}}_{{\scriptscriptstyle \text{\velo}}}}}

\def\lhcb {\mbox{LHCb}\xspace}

\def\alice  {\mbox{ALICE}\xspace}

\def\lhc    {\mbox{LHC}\xspace}

\def\velo   {VELO\xspace}

\def\MagUp {\mbox{\em Mag\kern -0.05em Up}\xspace}

\ifthenelse{\boolean{uprightparticles}}%
{

 \def\PDelta      {\ensuremath{\Delta}\xspace}                 
 \def\PXi      {\ensuremath{\Xi}\xspace}                 
 \def\PLambda      {\ensuremath{\Lambda}\xspace}                 
 \def\PSigma      {\ensuremath{\Sigma}\xspace}                 
 \def\POmega      {\ensuremath{\Omega}\xspace}                 
 \def\PUpsilon      {\ensuremath{\Upsilon}\xspace}                 
 
 \def\PB      {\ensuremath{\mathrm{B}}\xspace}                 
                  
 \def\PD      {\ensuremath{\mathrm{D}}\xspace}

 \def\PK      {\ensuremath{\mathrm{K}}\xspace}

 \def\Pb      {\ensuremath{\mathrm{b}}\xspace}                 
 \def\Pc      {\ensuremath{\mathrm{c}}\xspace}

 \def\Pi      {\ensuremath{\mathrm{i}}\xspace}

}
{

 \mathchardef\PDelta="7101
 \mathchardef\PXi="7104
 \mathchardef\PLambda="7103
 \mathchardef\PSigma="7106
 \mathchardef\POmega="710A
 \mathchardef\PUpsilon="7107
                  
 \def\PB      {\ensuremath{B}\xspace}                 
                  
 \def\PD      {\ensuremath{D}\xspace}

 \def\PK      {\ensuremath{K}\xspace}

 \def\Pb      {\ensuremath{b}\xspace}                 
 \def\Pc      {\ensuremath{c}\xspace}

 \def\Pi      {\ensuremath{i}\xspace}

}

\makeatletter
\ifcase \@ptsize \relax
  \newcommand{\miniscule}{\@setfontsize\miniscule{4}{5}}
\or
  \newcommand{\miniscule}{\@setfontsize\miniscule{5}{6}}
\or
  \newcommand{\miniscule}{\@setfontsize\miniscule{5}{6}}
\fi
\makeatother

\DeclareRobustCommand{\optbar}[1]{\shortstack{{\miniscule (\rule[.5ex]{1.25em}{.18mm})}
  \\ [-.7ex] $#1$}}

\def\cquark    {{\ensuremath{\Pc}}\xspace}

\def\bquark    {{\ensuremath{\Pb}}\xspace}

  \def\Kbar    {{\kern 0.2em\overline{\kern -0.2em \PK}{}}\xspace}

\def\KorKbar    {\kern 0.18em\optbar{\kern -0.18em K}{}\xspace}

  \def\Dbar    {{\kern 0.2em\overline{\kern -0.2em \PD}{}}\xspace}

\def\DorDbar    {\kern 0.18em\optbar{\kern -0.18em D}{}\xspace}

\def\Bbar    {{\ensuremath{\kern 0.18em\overline{\kern -0.18em \PB}{}}}\xspace}

\def\BorBbar    {\kern 0.18em\optbar{\kern -0.18em B}{}\xspace}

  \def\Y#1S{\ensuremath{\PUpsilon{(#1S)}}\xspace}

\def\Lbar        {{\ensuremath{\kern 0.1em\overline{\kern -0.1em\PLambda}}}\xspace}
\def\LorLbar    {\kern 0.18em\optbar{\kern -0.18em \PLambda}{}\xspace}

\def\AT#1     {\ensuremath{A_{\mathrm{T}}^{#1}}\xspace}           

\def\C#1      {\ensuremath{\mathcal{C}_{#1}}\xspace}                       
\def\Cp#1     {\ensuremath{\mathcal{C}_{#1}^{'}}\xspace}                    
\def\Ceff#1   {\ensuremath{\mathcal{C}_{#1}^{\mathrm{(eff)}}}\xspace}        
\def\Cpeff#1  {\ensuremath{\mathcal{C}_{#1}^{'\mathrm{(eff)}}}\xspace}       
\def\Ope#1    {\ensuremath{\mathcal{O}_{#1}}\xspace}                       
\def\Opep#1   {\ensuremath{\mathcal{O}_{#1}^{'}}\xspace}                    

\newcommand{\tev}{\ifthenelse{\boolean{inbibliography}}{\ensuremath{~T\kern -0.05em eV}\xspace}{\ensuremath{\mathrm{\,Te\kern -0.1em V}}}\xspace}
\newcommand{\gev}{\ensuremath{\mathrm{\,Ge\kern -0.1em V}}\xspace}
\newcommand{\mev}{\ensuremath{\mathrm{\,Me\kern -0.1em V}}\xspace}
\newcommand{\kev}{\ensuremath{\mathrm{\,ke\kern -0.1em V}}\xspace}
\newcommand{\ev}{\ensuremath{\mathrm{\,e\kern -0.1em V}}\xspace}
\newcommand{\gevc}{\ensuremath{{\mathrm{\,Ge\kern -0.1em V\!/}c}}\xspace}
\newcommand{\mevc}{\ensuremath{{\mathrm{\,Me\kern -0.1em V\!/}c}}\xspace}
\newcommand{\gevcc}{\ensuremath{{\mathrm{\,Ge\kern -0.1em V\!/}c^2}}\xspace}
\newcommand{\gevgevcccc}{\ensuremath{{\mathrm{\,Ge\kern -0.1em V^2\!/}c^4}}\xspace}
\newcommand{\mevcc}{\ensuremath{{\mathrm{\,Me\kern -0.1em V\!/}c^2}}\xspace}

\def\cm   {\ensuremath{\mathrm{ \,cm}}\xspace}

\def\mm   {\ensuremath{\mathrm{ \,mm}}\xspace}

\def\mum  {\ensuremath{{\,\upmu\mathrm{m}}}\xspace}

\def\invnb {\ensuremath{\mbox{\,nb}^{-1}}\xspace}

\def\gsim{{~\raise.15em\hbox{$>$}\kern-.85em
          \lower.35em\hbox{$\sim$}~}\xspace}
\def\lsim{{~\raise.15em\hbox{$<$}\kern-.85em
          \lower.35em\hbox{$\sim$}~}\xspace}

\def\ptot       {\mbox{$p$}\xspace}
\def\pt         {\mbox{$p_{\mathrm{ T}}$}\xspace}

\def\evtgen     {\mbox{\textsc{EvtGen}}\xspace}

\def\geant      {\mbox{\textsc{Geant4}}\xspace}

\def\pythia     {\mbox{\textsc{Pythia}}\xspace}

\def\tell1  {TELL1\xspace}
\def\ukl1   {UKL1\xspace}

\newcommand{\eg}{\mbox{\itshape e.g.}\xspace}

\usepackage{cite} 
\usepackage{mciteplus}

\begin{document}

\renewcommand{\thefootnote}{\fnsymbol{footnote}}
\setcounter{footnote}{1}

\begin{titlepage}
\pagenumbering{roman}

\vspace*{-1.5cm}
\centerline{\large EUROPEAN ORGANIZATION FOR NUCLEAR RESEARCH (CERN)}
\vspace*{1.5cm}
\noindent
\begin{tabular*}{\linewidth}{lc@{\extracolsep{\fill}}r@{\extracolsep{0pt}}}
\ifthenelse{\boolean{pdflatex}}
{\vspace*{-2.7cm}\mbox{\!\!\!\includegraphics[width=.14\textwidth]{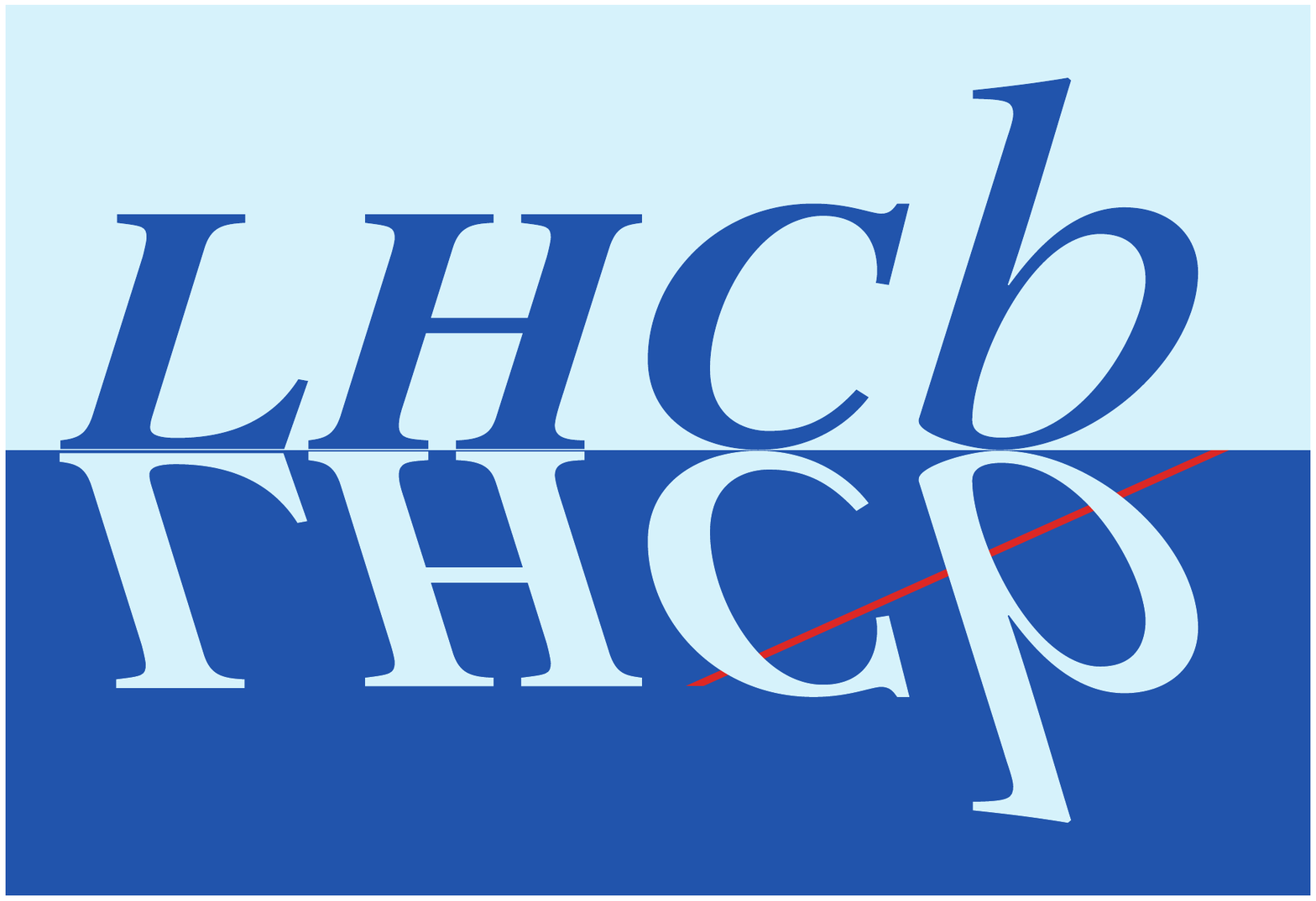}} & &}%
{\vspace*{-1.2cm}\mbox{\!\!\!\includegraphics[width=.12\textwidth]{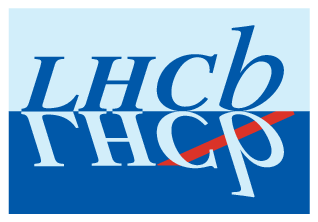}} & &}%
\\
 & & CERN-PH-EP-2015-308 \\  
 & & LHCb-PAPER-2015-040 \\  
 & & October 6, 2016 \\ 
 & & \\
\end{tabular*}

\vspace*{1.0cm}

{\normalfont\bfseries\boldmath\huge
\begin{center}
  Measurements of long-range near-side angular correlations in $\sqrtnn=5\tev$ proton-lead collisions in the forward region
\end{center}
}

\vspace*{1.0cm}

\begin{center}
The LHCb collaboration\footnote{Authors are listed at the end of this letter.}
\end{center}

\vspace{\fill}

\begin{abstract}
  \noindent
  Two-particle angular correlations are studied in proton-lead collisions at a nucleon-nucleon centre-of-mass energy of $\sqrtnn=5\tev$, collected with the \lhcb detector at the \lhc. 
  The analysis is based on data recorded in two beam configurations, in which either the direction of the proton or that of the lead ion is analysed.
  The correlations are measured in the laboratory system as a function of relative pseudorapidity, $\deta$, and relative azimuthal angle, $\dphi$, for events in different classes of event activity and for different bins of particle transverse momentum. 
  In high-activity events a long-range correlation on the near side, $\dphi \approx 0$, is observed in the pseudorapidity range $2.0<\eta<4.9$.   
  This measurement of long-range correlations on the near side in proton-lead collisions extends previous observations into the forward region up to $\eta=4.9$.
  The correlation increases with growing event activity and is found to be more pronounced in the direction of the lead beam. 
  However, the correlation in the direction of the lead and proton beams are found to be compatible when comparing events with similar absolute activity in the direction analysed.
\end{abstract}

\vspace*{1.0cm}

\begin{center}
  Published in Phys.~Lett.~B762 (2016) 473-483
\end{center}

\vspace{\fill}

{\footnotesize 
\centerline{\copyright~CERN on behalf of the \lhcb collaboration, licence \href{http://creativecommons.org/licenses/by/4.0/}{CC-BY-4.0}.}}
\vspace*{2mm}

\end{titlepage}


\newpage
\setcounter{page}{2}
\mbox{~}

\cleardoublepage

\renewcommand{\thefootnote}{\arabic{footnote}}
\setcounter{footnote}{0}



\pagestyle{plain} 
\setcounter{page}{1}
\pagenumbering{arabic}

\section{Introduction}
\label{sec:Introduction}

Studies of two-particle angular correlations are an important experimental method 
to investigate the dynamics of multi-particle production in QCD and to probe collective effects arising in the dense environment of a high-energy collision.
The highest particle densities and multiplicities reached in proton-proton (\pp) and proton-lead collisions (\pa) at the \lhc are of a similar size to those in non-central nucleus-nucleus (AA) collisions.
This motivates looking for signatures which were so far mainly studied in AA collisions.

Two-particle correlations are conveniently described by two-dimensional ($\deta,\dphi$)-correlation functions.
For pairs of prompt charged particles their separations in pseudorapidity, $\deta$, and in the azimuthal angle, $\dphi$, are measured in the laboratory system. 
Structures in the correlation function are classified into short-range ($|\deta|\lesssim2$) and long-range ($|\deta|\gtrsim2$) effects. 
On the near-side ($|\dphi|\approx0$) a short-range ``jet peak" at $\deta\approx0$ is the dominant structure, caused by the fact that in 
the fragmentation process the final-state particles are collimated around the initial parton. 
To balance the momentum, the peak is accompanied by a long-range structure on the away side ($|\dphi|\approx\pi$) caused by particles that are opposite in azimuthal angle.

Due to the different momentum fractions carried by the colliding partons and the resulting individual boosts, the away-side structure is not restricted in $\deta$, but elongated over a large range.
In complex heavy-ion collisions, these short- and long-range structures are modified as a result of the strongly interacting medium that is formed depending on the centrality of the collision.
Long-range correlations on the near- and away-side are observed, which are typically explained as being the result of a hydrodynamical flow of the deconfined medium~\cite{Ollitrault:1992bk}.
Measurements in very rare \pp collisions that have an extremely high particle multiplicity revealed a similar unexpected long-range correlation on the near side~\cite{Khachatryan:2010gv,Aad:2015gqa,Khachatryan:2015lva}.
This structure, often referred to as the near-side ``ridge'', has also been confirmed in high-multiplicity \pa collisions~\cite{CMS:2012qk,Abelev:2012ola,Aad:2012gla,Aad:2014lta,Adam:2015bka}, where it was found to be much more pronounced than in \pp collisions.

The theoretical interpretation of the mechanism responsible for the ridge in \pp and \pa is still under discussion.
Various models have been proposed such as gluon saturation in the framework of a colour-glass condensate~\cite{Dusling:2012cg,Dusling:2012wy,Dusling:2013oia,Kovchegov:2012nd} or the hydrodynamic evolution of a high density
partonic medium~\cite{Bzdak:2013zma}, 
multiparton interactions~\cite{Alderweireldt:2012kt,Strikman:2011cx,Ryskin:2011qh}, 
jet-medium interactions~\cite{Hwa:2010fj,Wong:2011qr}, and collective effects~\cite{Avsar:2010rf,Werner:2010ss,Bozek:2011if,Bozek:2012gr,Shuryak:2013ke} induced by the formation and expansion of 
a high-density system possibly produced in these collisions.
Analyses that have seen the near-side ridge at the \lhc have been performed in the central rapidity region, probing ranges up to $|\eta|=2.5$.
In a recent analysis~\cite{Adam:2015bka} larger pseudorapidities were also accessed in measurements of muon-hadron correlations between the forward ($2.5<|\eta|<4.0$) and the central ($|\eta|<1.0$) region. 
For the present measurement the forward acceptance of the \lhcb detector, unique among the \lhc experiments, is used to study the ridge phenomenon in \pa collisions.
Proton-lead collisions are analysed in the range of $2.0<\eta<4.9$ and in the directions of the proton and the lead beams separately.
Confirmation of the ridge correlation at large pseudorapidities and comparison of its magnitude for the two beam directions provide new input to the theoretical understanding of the underlying mechanisms.

\section{Experimental setup}
\label{sec:Setup}

The analysis is based on data collected with the \lhcb detector during the proton-lead data-taking period in 2013.
The \lhc provided \pa collisions at a nucleon-nucleon centre-of-mass energy of $\sqrtnn=5\tev$, corresponding to a proton beam energy of $4\tev$ and a lead beam energy of $1.58\tev$ per nucleon.
Due to this asymmetric beam configuration, there is a relative boost between the rapidity in the \lhcb laboratory frame, $y_{\text{lab}}$, and $y$ in the nucleon-nucleon centre-of-mass frame, corresponding to a shift of $0.47$ units.

The \lhcb detector~\cite{Alves:2008zz,LHCb-DP-2014-002} is a single-arm forward
spectrometer covering the \mbox{pseudorapidity} range $2<\eta <5$ in the laboratory frame.
Depending on the direction of the proton and the lead beam, two different configurations are distinguished.
In the \textit{forward} configuration the proton beam points to positive rapidity, into the \lhcb spectrometer, and the recorded collisions are referred to as \ppb.
The opposite \textit{backward} configuration, in which the lead beam points to positive rapidity, is referred to as \pbp.
The measurement is performed in the \lhcb laboratory frame, probing rapidities $y$ in the nucleon-nucleon centre-of-mass frame of $1.5<y<4.4$ in the \ppb configuration and $-5.4<y<-2.5$ in the \pbp configuration.
The data used for this analysis correspond to an integrated luminosity of 0.46\invnb in the \ppb configuration and 0.30\invnb for the \pbp configuration.

The \lhcb detector, designed for the study of particles containing \bquark or \cquark
quarks, includes a high-precision tracking system
consisting of a silicon-strip vertex detector (\velo) surrounding the 
interaction region, a large-area silicon-strip detector located
upstream of a dipole magnet with a bending power of about
$4{\rm\,Tm}$, and three stations of silicon-strip detectors and straw
drift tubes placed downstream of the magnet.
The polarity of the dipole magnet was reversed once for each configuration 
to average over small asymmetries in the detection of charged particles. 
The tracking system provides a measurement of momentum of charged particles with
a relative uncertainty that varies from 0.5\% at low momentum to 1.0\% at 200\gevc.
Different types of charged hadrons are distinguished using information 
from two ring-imaging Cherenkov detectors. 
Photons, electrons and hadrons are identified by a calorimeter system consisting of
scintillating-pad and preshower detectors, an electromagnetic
calorimeter and a hadronic calorimeter. Muons are identified by a
system composed of alternating layers of iron and multiwire
proportional chambers.
The online event selection is performed by a trigger, 
which consists of a hardware stage, based on information from the calorimeter and muon
systems, followed by a software stage, which applies a full event
reconstruction.
During data taking of \pa collisions, an activity trigger in the hardware stage accepted non-empty beam bunch crossings, 
and the software stage accepted events with at least one reconstructed track in the \velo.

\section{Data selection and corrections}
\label{sec:Selection}

Monte Carlo simulations are used to evaluate the efficiency of the following selections and to estimate the remaining contamination in the selected track sample.
Proton-lead collisions in \ppb and \pbp configurations are simulated using the \hijing generator~\cite{Wang:1991hta} in version 1.383bs.2.
As a cross-check, proton-proton collisions at a centre-of-mass energy of $8\tev$ are simulated using \pythia~\cite{Sjostrand:2007gs,*Sjostrand:2006za} in a special \lhcb configuration~\cite{LHCb-PROC-2010-056} and with a high average interaction rate (large pile-up) to reproduce the larger particle multiplicity in proton-lead collisions.
Particle decays are simulated by \evtgen~\cite{Lange:2001uf}. 
The interaction of the generated particles with the detector, and its response, are implemented using the \geant toolkit~\cite{Allison:2006ve, *Agostinelli:2002hh} as described in Ref.~\cite{LHCb-PROC-2011-006}.

The measurements are based on proton-lead collisions that are dominated by single interactions; 
fewer than $2\%$ of the bunch crossings have more than one interaction.
Each event is required to have exactly one reconstructed primary vertex containing at least five tracks.
Beam-related background interactions are suppressed by requiring the position of the reconstructed primary vertex to be within $\pm3$ standard deviations around the mean interaction point, separately for each coordinate.
The mean value and the width of this luminous region are determined separately from a Gaussian fit to the distribution of reconstructed primary vertices of each data sample.
Depending on the polarity of the magnetic field and the resulting beam optics, the size of the standard deviation along the beam axis is approximately 40\mm or 60\mm, while in the transverse direction it is around 30\mum. 
While \pa interactions are most likely produced in this region, beam-related background extends further along the beam line. 
Beam gas events or interactions with detector material can produce a very high number of particles; 
however, in such cases the total energy deposit in the calorimeter is much smaller than that of typical \pa collisions. 
Events with too small a ratio of the number of clusters in the electromagnetic calorimeter to that in the \velo are rejected; 
individual lower bounds are defined for collisions in the \ppb and \pbp configuration using simulation.

The angular correlations are determined for charged particles that are directly produced in the \pa interaction.
The measurement is based on tracks traversing the full tracking system, which restricts charged particles in pseudorapidity to $2.0<\eta<4.9$. 
In addition, particles are required to have a transverse momentum $\pt>0.15\gevc$ and a total momentum $\ptot>2\gevc$.
Reconstruction artefacts, such as fake tracks, are suppressed using a multivariate classifier.
The remaining average fraction of fake tracks is of the order of $7\%$ and $12\%$ in the \ppb and \pbp samples, respectively. 
The probability of reconstructing fake tracks increases with the number of hits in the tracking detectors.
Thus, the difference between the data samples is due to the higher average particle and hit multiplicity that is present in the direction of the lead remnant.
To select primary tracks originating directly from the \pa collision the impact parameter of each track with respect to the reconstructed primary vertex must not exceed $1.2\mm$, after which the fraction of remaining tracks from secondary particles is estimated to be less than $3.5\%$.

The inefficiency in finding charged particles arises from two effects:
limited detector acceptance in the range of $2.0<\eta<4.9$, and limitations of the track reconstruction. 
For particles fulfilling the kinematic requirements, the acceptance describes the fraction that reach the end of the downstream
tracking stations and is about $70\%$ on average.
In contrast, the track reconstruction efficiency varies from $96\%$ for low-multiplicity events to $60\%$ for events with the highest measured multiplicity.

After applying the selection requirements, the remaining probabilities of selecting fake tracks, $\mathcal{P}_{\text{fake}}$, and secondary particles, $\mathcal{P}_{\text{sec}}$, as well as the efficiencies related to the detector acceptance, $\epsilon_{\text{acc}}$, and the track reconstruction, $\epsilon_{\text{tr}}$, are estimated in simulation as a function of the angular variables $\eta$ and $\phi$, the transverse momentum \pt, and the hit-multiplicity in the \velo, \nhitvelo.
Each reconstructed track is assigned a weight, $\omega$, that accounts for these effects:
\begin{equation}
 \omega(\eta,\phi,\pt,\nhitvelo)= (1-\mathcal{P}_{\text{fake}}-\mathcal{P}_{\text{sec}})/(\epsilon_{\text{acc}} \cdot \epsilon_{\text{tr}}).
\label{eq:weights}
\end{equation}

\section{Activity classes and data samples}
\label{sec:ActivityClasses}

Two-particle correlations show a strong dependence on the number of particles produced within a collision.
The hit multiplicity in the \velo is proportional to this global event property. 
With its coverage in pseudorapidity ranging from $1.9<\eta<4.9$ in the forward direction and $-2.5<\eta<-2.0$ in the backward direction, the \velo can probe the total number of charged particles per event more comprehensively than other sub-detectors of \lhcb.

The analysis presented in this paper is based on a subset of the total data set recorded during the 2013 \pa running period. 
The \ppb and \pbp minimum bias samples each consist of about $1.1\times 10^{8}$ events which are randomly selected from the about $10$ times larger full sample. 
The high-multiplicity samples contain all recorded events with at least $2200$ hits in the \velo and amount to about $1.1\times 10^{8}$ events in \pbp and $1.3\times 10^{8}$ events in \pbp collisions.

\begin{figure}[!tb]  
  \begin{center}
    \includegraphics[width=0.49\linewidth]{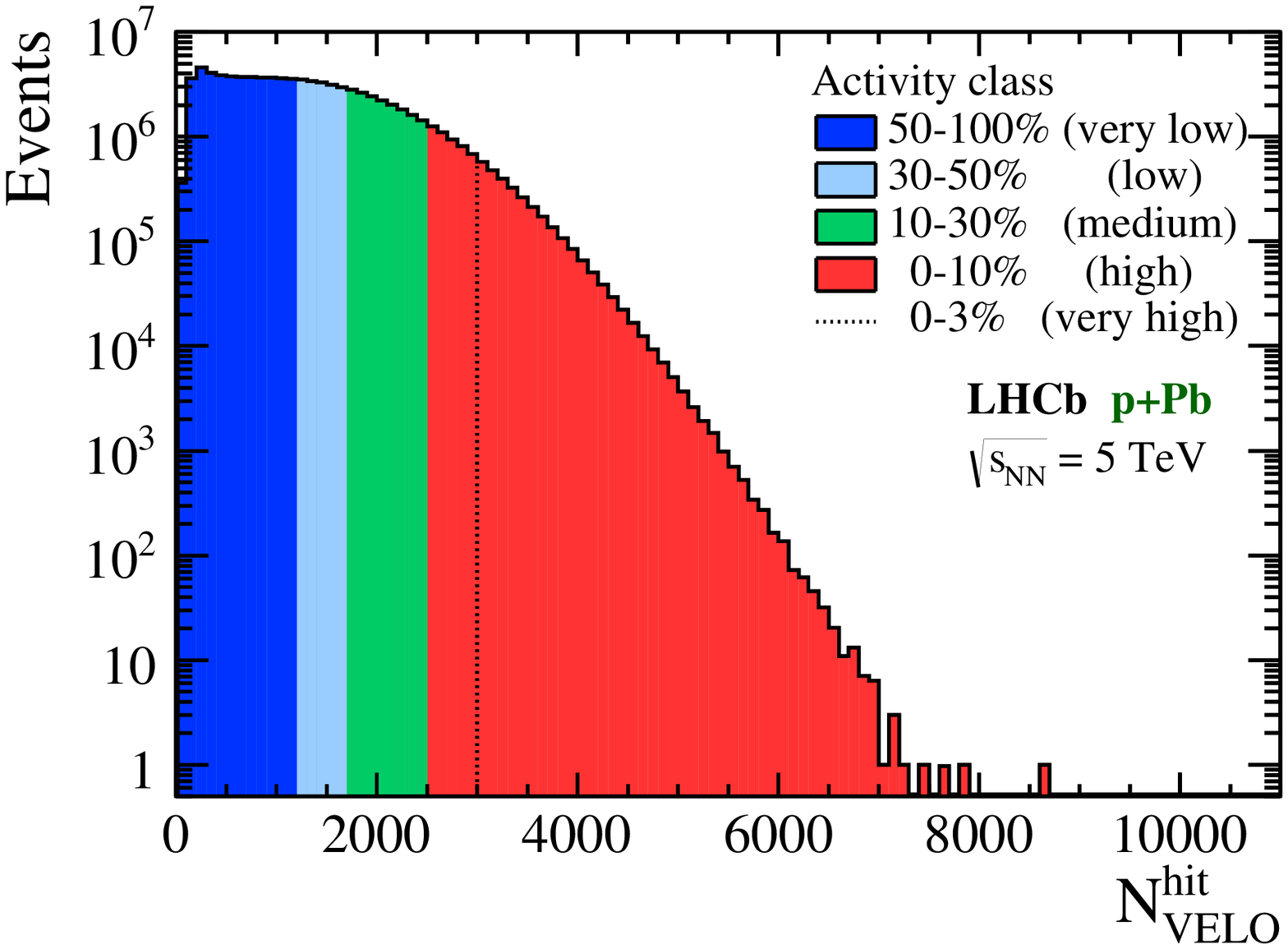}
    \includegraphics[width=0.49\linewidth]{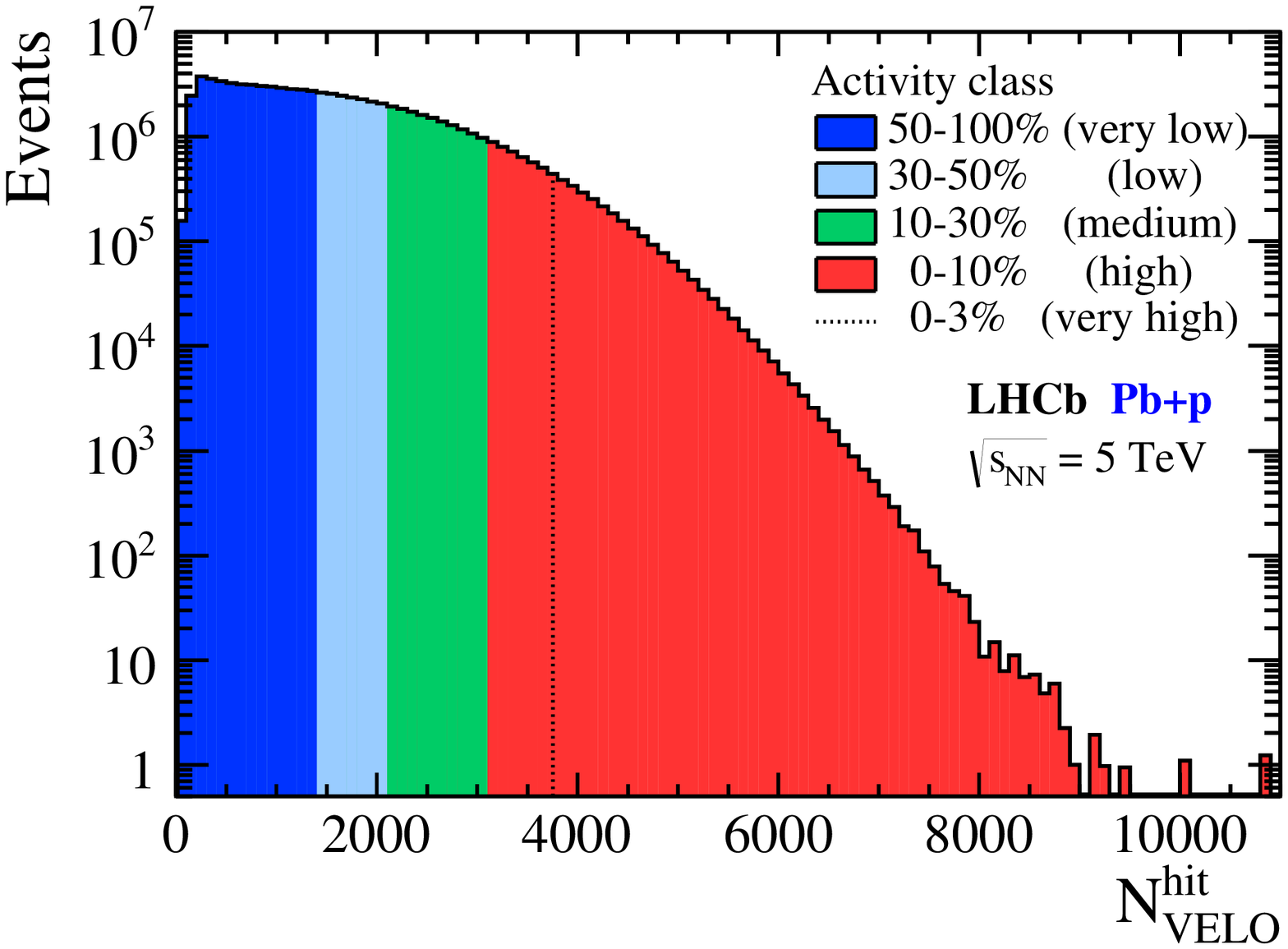}
  \end{center}
  \caption{
    \small
    Hit-multiplicity distribution in the \velo for selected events of the minimum-bias samples in the (left) \ppb and (right) \pbp configurations.
    The activity classes are defined as fractions of the full distribution, as indicated by colours (shades).
    The $0-3\%$ class is a sub-sample of the $0-10\%$ class.
  }
  \label{fig:relActivityClasses}
\end{figure}

Five event-activity classes are defined as fractions of the hit-multiplicity distributions of the minimum-bias samples, as indicated in 
Fig.~\ref{fig:relActivityClasses}.
Since collisions recorded in the \pbp configuration reach larger hit-multiplicities compared to those in the \ppb configuration, the relative classes are defined separately for each configuration.
The $50-100\%$ class contains approximately the $50\%$ of events with the lowest event activities, followed by the $30-50\%$ and $10-30\%$ classes representing medium-activity events, and the $0-10\%$ and $0-3\%$ classes of high-activity events. 
The ranges defining the activity classes are listed in Table~\ref{tab:RelActivityClasses}.
For each class, average numbers of charged particles, $\langle N_{ch} \rangle_{\text{MC}}$, are quoted for illustration, based on the \hijing event generator.

\begin{table}[b] 
\caption{\small
  Relative event-activity classes defined by the \velo-hit multiplicity, \nhitvelo, of an event.
  The classes are defined as fractions of the \nhitvelo distribution for minimum-bias recorded events in the \ppb or \pbp configuration. 
  The $0-3\%$ class is a sub-sample of the $0-10\%$ class.
  For illustration purposes the average number, $\langle N_{ch} \rangle_{\text{MC}}$, of prompt charged particles with $\ptot>2\gevc$, $\pt>0.15\gevc$ and $2.0<\eta<4.9$ is listed for events simulated with the \hijing event generator.
  Statistical uncertainties are negligible.
  }
\begin{center}  
  \begin{tabular}{l|cc|cc}
    \hline
    Relative & \multicolumn{2}{c|}{\ppb} &\multicolumn{2}{c}{\pbp} \bigstrut[tb] \\
    activity class & range \nhitvelo & $\langle N_{ch} \rangle_{\text{MC}}$ & range \nhitvelo & $\langle N_{ch} \rangle_{\text{MC}}$ \\  
    \hline
    $50-100\%$ very low 		& $\phantom{000}0-1200$ 	& 18.9 & $\phantom{000}0-1350$ \bigstrut[t] 	& 29.2\\
    $30-50\%$ \phantom{0}low		& $1200-1700$ 			& 30.0 & $1350-2000$ 				& 47.4 \\
    $10-30\%$ \phantom{0}medium 	& $1700-2400$ 			& 42.8 & $2000-3000$ 				& 70.9 \\
    $\phantom{0}0-10\%$  \phantom{0}high  	& $2400-\text{max} $ 	& 63.6 & $3000-\text{max}$ 			& 106.7 \\
    $\phantom{0}0-3\%$   \phantom{00}very high& $3000-\text{max} $ 	& 73.7 & $3800-\text{max}$ 			& 126.4 \\
    \hline
  \end{tabular}
\end{center}
\label{tab:RelActivityClasses}
\end{table}

The long-range correlations in the direction of the fragmenting proton (\ppb configuration) and the direction of the fragmenting lead ion (\pbp configuration) are compared for classes of the same absolute activity in the pseudorapidity range of $2.0<\eta<4.9$. 
Here a proper assignment of equivalent activity classes needs to take into account the fact that the \velo acceptance is larger than the pseudorapidity interval of interest. 
Assuming a linear relation between the total number of \velo hits and the number of tracks in the range $2.0<\eta<4.9$, one finds that $N$ \velo hits in the \pbp configuration correspond to $N/(0.77\pm0.08)$ \velo hits in the \pbp case. 
The uncertainty in the scaling factor accounts for deviations from perfect linearity in the data that are not reproduced in the simulation, and is propagated into the systematic uncertainties of the results.
Five common absolute activity classes, labelled I -- V, are defined in the high-activity region and are listed in Table~\ref{tab:AbsActivityBins} with the corresponding average numbers of charged particles from simulation.
The quoted uncertainties in the \ppb sample are related to the systematic uncertainty of the scaling factor.

The analysis is repeated using an alternative event-activity classification, based on the multiplicity of selected tracks in the range $2.0<\eta<4.9$.
In analogy to the nominal approach using the \velo-hit multiplicity, the same fractions of the full distribution are used to define relative activity classes for both beam configurations. 
Similarly, five common activity bins for the \ppb and \pbp samples are defined in the intermediate to high-activity classes. 
The results are found to be independent of the definition of the activity classes.

\begin{table}[t] 
\caption{\small
    Common absolute activity bins for the \ppb and \pbp samples. 
    The activity of \ppb events is scaled to match the same activity ranges of \pbp events, as explained in the text.
    For illustration purposes the average number, $\langle N_{ch} \rangle_{\text{MC}}$, of prompt charged particles with $\ptot>2\gevc$, $\pt>0.15\gevc$ and $2.0<\eta<4.9$ 
    is listed for events simulated with the \hijing event generator.
    The uncertainties are due to the scaling factor of $0.77 \pm 0.08$. 
    Statistical uncertainties are negligible.
    }
\begin{center}  
  \begin{tabular}{l|c|c|c}
    \hline
    Common absolute & \nhitvelo-range & \ppb & \pbp \bigstrut[t] \\
    activity bin    & in \pbp scale & $\langle N_{ch} \rangle_{\text{MC}} $ & $\langle N_{ch} \rangle_{\text{MC}} $ \\  
    \hline
    Bin I\phantom{II} 	& $2200-2400$ & $62.8 \pm 6.6$ & $64.4$ \\
    Bin II\phantom{I} 	& $2400-2600$ & $68.4 \pm 7.1$ & $67.0$ \\
    Bin III		& $2600-2800$ & $73.7 \pm 7.6$ & $76.4$ \\
    Bin IV 		& $2800-3000$ & $79.2 \pm 7.9$ & $82.4$ \\
    Bin V\phantom{I}	& $3000-3500$ & $86.7 \pm 8.2$ & $92.9$ \\
    \hline
  \end{tabular}
\end{center}
\label{tab:AbsActivityBins}
\end{table}

\section{Analysis method}
\label{sec:Analysis} 

Two-particle correlations are measured separately for events in each activity class. 
The track sample containing the selected candidates of primary charged particles is divided into three \pt intervals: $0.15-1.0\gevc$, $1.0-2.0\gevc$ and $2.0-3.0\gevc$.
For each event, all candidates within a given \pt interval are identified as \textit{trigger} particles. 
By selecting a trigger particle all remaining candidates within the same interval compose the group of \textit{associated} particles.
Particle pairs are formed by combining every trigger particle with each associated particle. 
Due to the symmetry around the origin, differences in azimuthal angle $\dphi$ are taken in the range $[0, \pi]$ and as absolute values in $\deta$. 
For visualisation purposes plots are symmetrized.
The two-particle correlation function is composed of a signal part $S(\deta,\dphi)$, a background part $B(\deta,\dphi)$, and a normalization factor $B(0,0)$. 
The total function is defined as the associated yield per trigger particle, given by
\begin{equation}
 \frac{1}{N_{\text{trig}}}\frac{\text{d}^{2}N_{\text{pair}}}{\text{d}\Delta\eta\,\text{d}\Delta\phi} = \frac{S(\Delta\eta,\Delta\phi)}{B(\Delta\eta,\Delta\phi)} \times B(0,0),
\label{eq:correlation}
\end{equation}
where $N_{\text{pair}}$ is the number of particle pairs found in a ($\Delta\eta,\Delta\phi$) bin. 
The number of trigger particles within a given \pt interval and activity class is denoted by $N_{\text{trig}}$.
The signal distribution $S(\deta,\dphi)$ describes the associated yield per trigger particle for particle pairs, $N_{\text{same}}$, formed from the same event, and is defined as
\begin{equation}
 S(\deta,\dphi) = \frac{1}{N_{\text{trig}}}\frac{\text{d}^{2}N_{\text{same}}}{\text{d}\Delta\eta\,\text{d}\Delta\phi}.
\end{equation}
Following the approach in Ref.~\cite{Abelev:2012ola}, the sum over the events is performed separately for $N_{\text{trig}}$ and for $\text{d}^{2}N_{\text{same}}/\text{d}\Delta\eta\,\text{d}\Delta\phi$ before the ratio is calculated. 
The background distribution $B(\deta,\dphi)$ is defined for particle pairs of mixed events,
\begin{equation}
 B(\deta,\dphi) = \frac{\text{d}^{2}N_{\text{mix}}}{\text{d}\Delta\eta\,\text{d}\Delta\phi},
\end{equation}
and describes the yield of uncorrelated particles. 
The $N_{\text{mix}}$ pairs are constructed by combining all trigger particles of an event with the associated particles of five different random events in the same activity class, whose 
vertex positions in the beam direction are within $2\cm$ of the original event. 
As a result, effects due to the detector occupancy, acceptance and material are accounted for by dividing the signal by the background distribution, where the latter is normalised to unity around the origin.
The factor $B(0,0)$ describes the associated yield for particles of a pair travelling in approximately the same direction and thus having the maximum pair acceptance.

All trigger and associated particles in the signal and background distributions are weighted with the correction factors $\omega$ described in Section~\ref{sec:Selection}.
Furthermore, alternative correction factors determined from the large pile-up \pp simulation using \pythia are applied to evaluate systematic uncertainties.
The resulting associated correlation yields agree within $3\%$ with the nominal results. 
To estimate the influence of the track selection, the correction factors are also determined with a maximum impact parameter relaxed to twice the nominal value, and the value of the multivariate classifier used to suppress fake tracks is varied by $\pm5\%$. 
The resulting different correction factors are applied to the measurements which are then compared to the nominal corrected results.
The difference due to the different prompt selection is negligible, while the alternative fake track suppression results in a maximum variation of $3\%$. 
Typical variations are much smaller. 
The effect on the final results, obtained after subtracting a global offset, is negligible.

\section{Results}
\label{sec:Results}

\begin{figure}[tb]
  \begin{center}
    \includegraphics[width=0.49\linewidth]{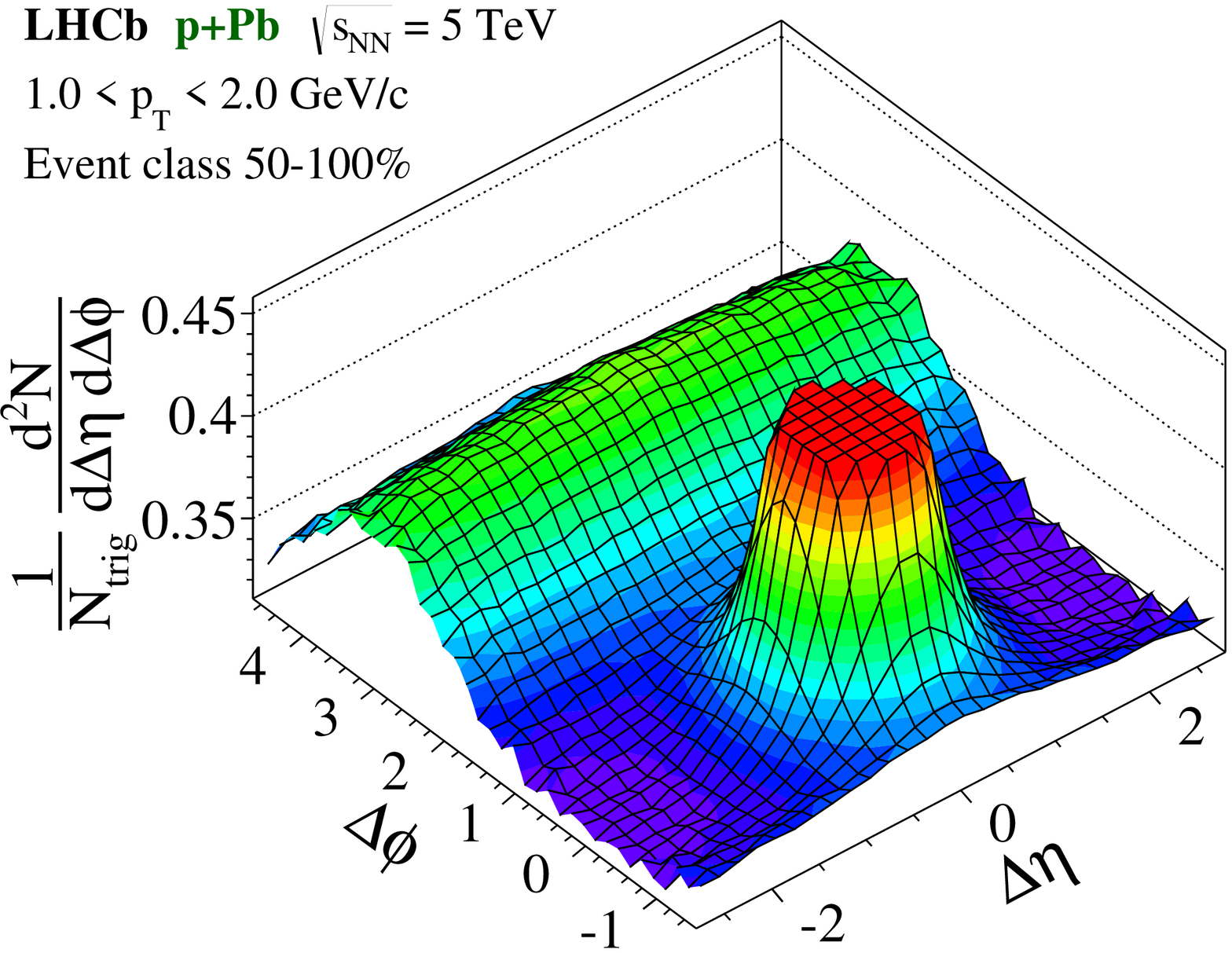}
    \includegraphics[width=0.49\linewidth]{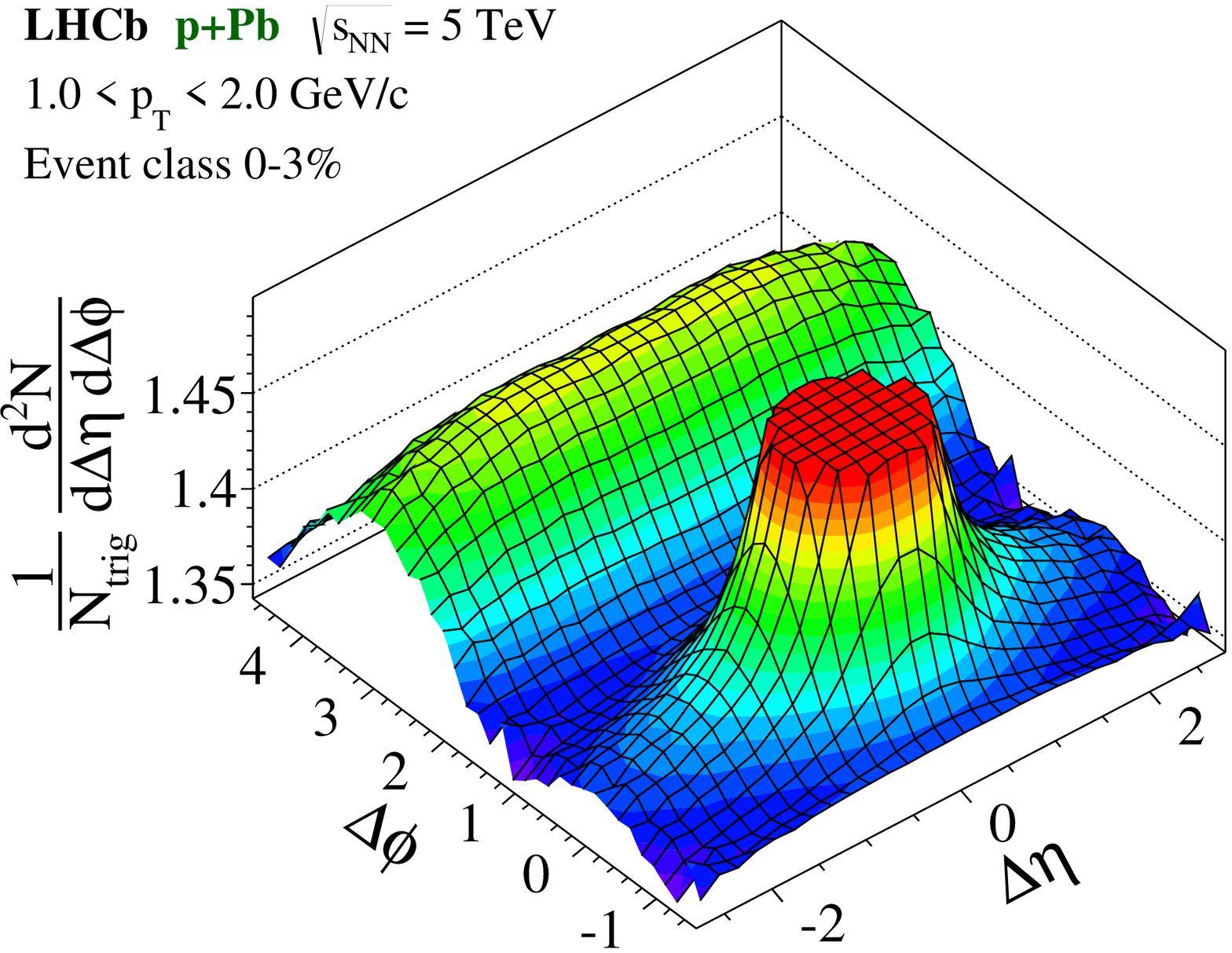}
  \end{center}
  \caption{\small
  Two-particle correlation functions for events recorded in the \ppb configuration, showing the (left) low and (right) high event-activity classes.
  The analysed pairs of prompt charged particles are selected in a \pt range of $1-2\gevc$.
  The near-side peak around $\deta=\dphi=0$ is truncated in the histograms.
  }
  \label{fig:cor_pA_pt1}
\end{figure}

\begin{figure}[tb]
  \begin{center}
    \includegraphics[width=0.49\linewidth]{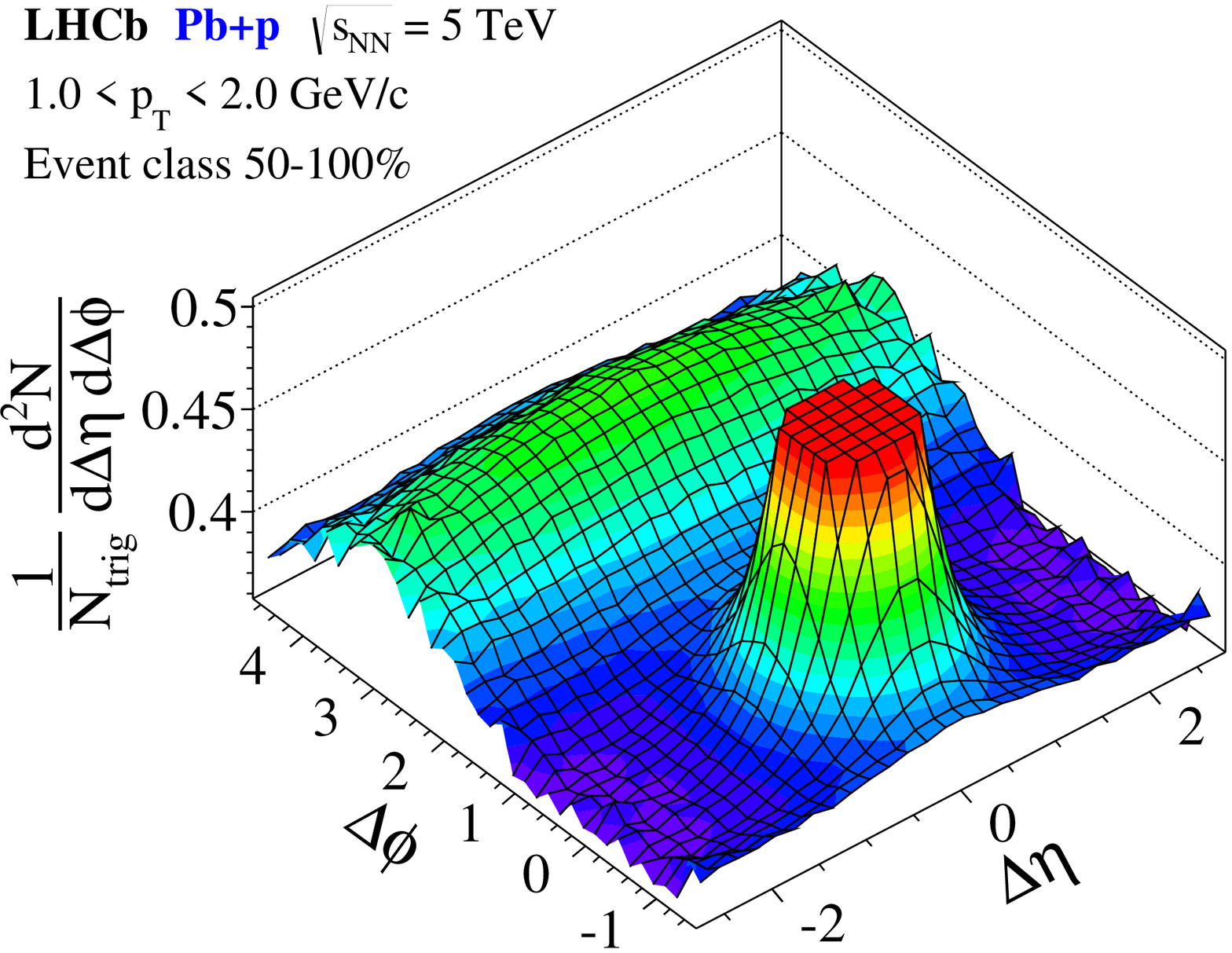}
    \includegraphics[width=0.49\linewidth]{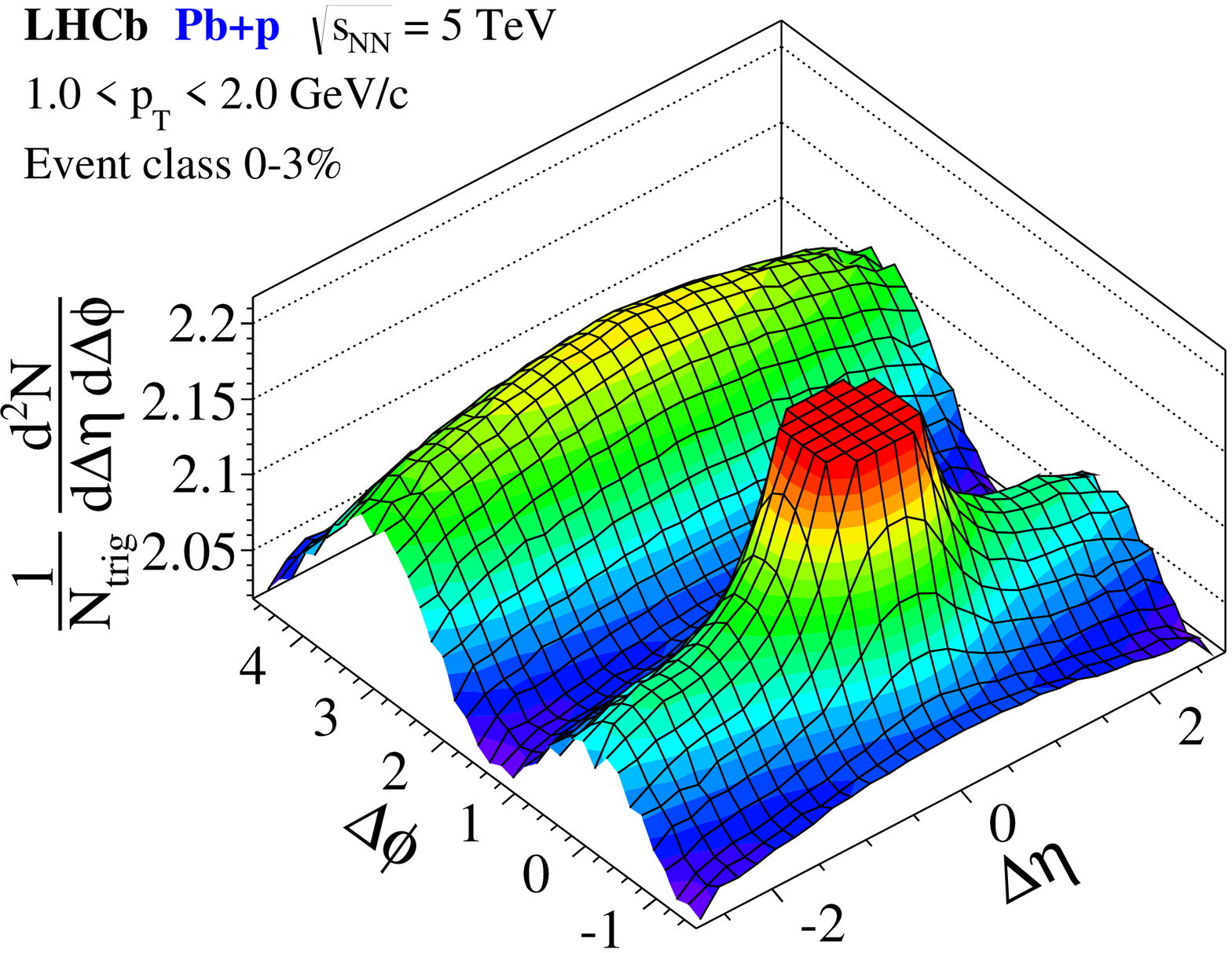}
  \end{center}
  \caption{\small
  Two-particle correlation functions for events recorded in the \pbp configuration, showing the (left) low and (right) high event-activity classes.
  The analysed pairs of prompt charged particles are selected in a \pt range of $1-2\gevc$.
  The near-side peak around $(\deta=\dphi=0)$ is truncated in the histograms.
  }
  \label{fig:cor_Ap_pt1}
\end{figure}

Two-particle correlation functions for events recorded in the \ppb configuration are presented in Fig.~\ref{fig:cor_pA_pt1}.
The correlation for particles with $1<\pt<2\gevc$ is shown for events of the $50-100\%$ and $0-3\%$ class, representing low and very-high event activities, respectively.
Both histograms are dominated by the jet peak around $\deta\approx\dphi\approx0$ which is due to correlations of particles originating from the same jet-like objects and thus being boosted closely together.
For better visualization of additional structures, in all 2D-histograms the jet peak is truncated.
The second prominent feature is visible on the away-side ($\dphi\approx\pi$) over a long range in $\deta$ and combines jet and (potential) ridge contributions.
The event sample with very high event activity (Fig.~\ref{fig:cor_pA_pt1}, right) shows an additional, less pronounced, long-range structure centred at $\dphi=0$, which is not present in the corresponding low-activity sample.
The structure, often referred to as the near-side ridge, is elongated over the full measured $\deta$ range of $2.9$ units.
This observation of the ridge for particles produced in proton-lead collisions at forward rapidities, $2.0<\eta<4.9$, extends previous  measurements at the \lhc.

Two-particle correlations for events recorded in the \pbp configuration are shown in Fig.~\ref{fig:cor_Ap_pt1}, for particle pairs with $1<\pt<2\gevc$.
The $50-100\%$ and $0-3\%$ activity classes in the \pbp sample exhibit the same correlation structures as the corresponding classes in the \ppb sample.
While the shape and magnitude of the jet peak and the away-side ridge appear to be of similar sizes in both beam configurations, the near-side ridge is more pronounced for particles in the direction of the lead beam.
For the $3\%$ of events with the highest event activity, the near-side ridge in the \pbp sample is much more prominent than that in the \ppb sample. 

\begin{figure}[tb]
  \begin{center}
    \includegraphics[width=0.49\linewidth]{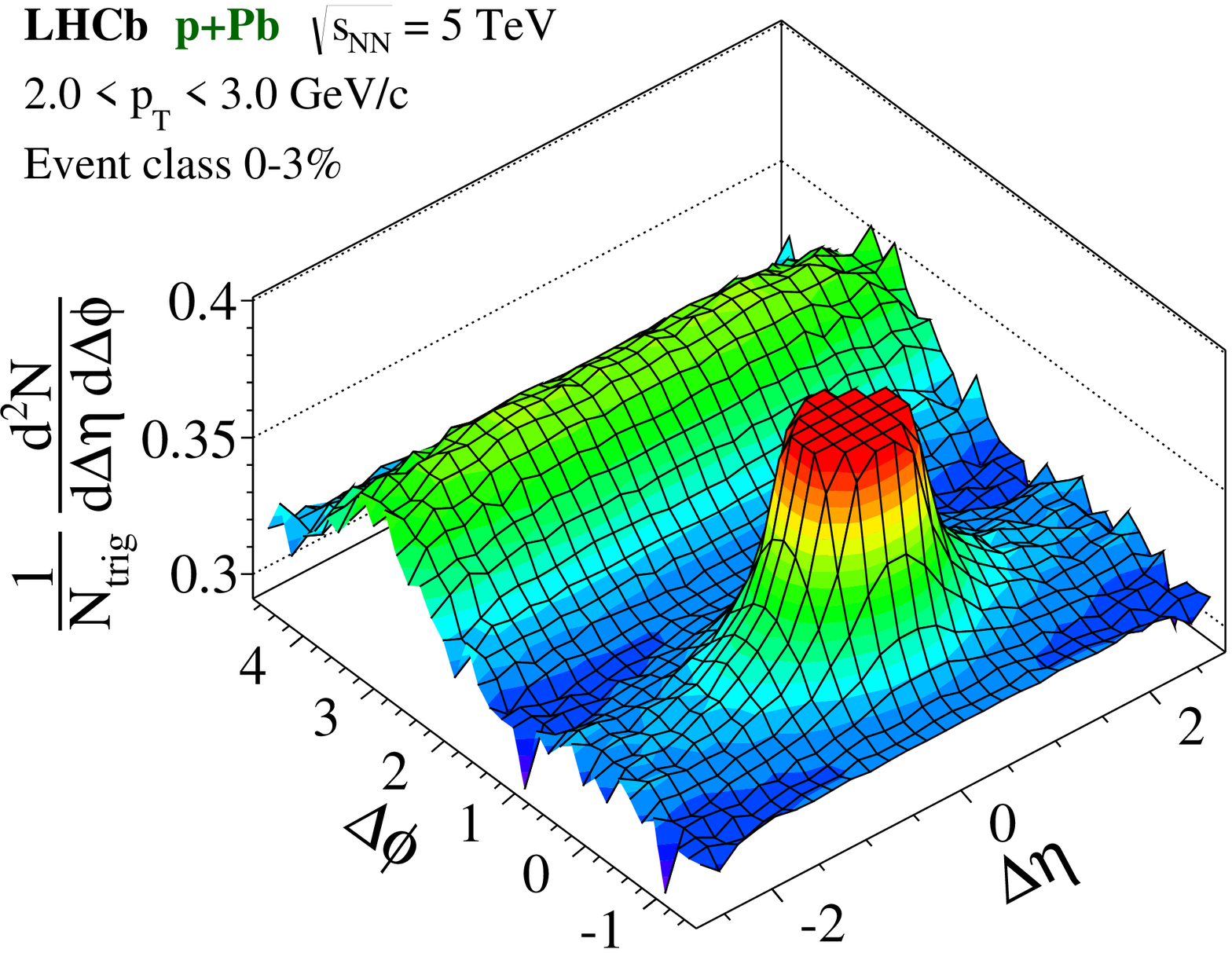}
    \includegraphics[width=0.49\linewidth]{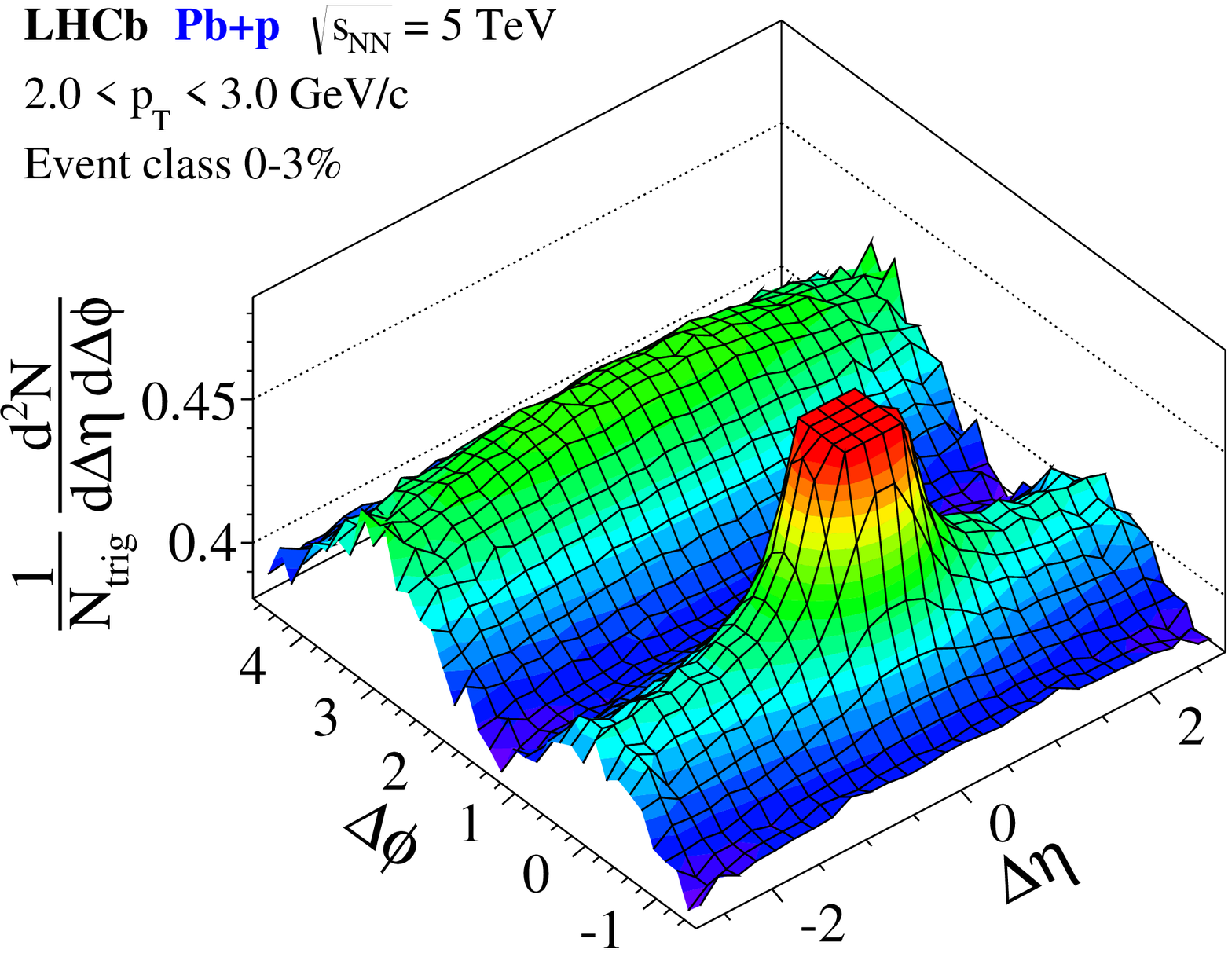}
  \end{center}
  \caption{\small
  Two-particle correlation functions for events recorded in the \ppb (left) and \pbp (right) configurations, showing the $0-3\%$ event-activity class.
  The analysed pairs of prompt charged particles are selected in a \pt range of $2-3\gevc$.
  The near-side peak around $(\deta=\dphi=0)$ is truncated in each histogram.
  }
  \label{fig:cor_pt2}
\end{figure}

Similar behaviour is found when analysing particle pairs with larger transverse momenta in the interval $2<\pt<3\gevc$.
In Fig.~\ref{fig:cor_pt2} the correlation functions in this \pt range are presented for the $3\%$ highest-activity events recorded in the \ppb and \pbp configurations. 
The near-side ridge is present in both samples; however in the \ppb sample it is only marginally visible while in the \pbp sample a strongly pronounced ridge is found.
The short-range jet peak in this higher \pt interval is more collimated compared to the $1-2\gevc$ interval, because of the higher average total momentum of the particles. 
As a result, the near-side ridge is visible towards $|\deta|$ values slightly below $2.0$ without being covered by the jet peak.

In order to study the evolution of the long-range correlations on the near and away sides in more detail, one-dimensional projections of the correlation function on $\dphi$ are calculated,
\begin{equation}
  Y(\dphi)\equiv\frac{1}{N_{\text{trig}}}\frac{\text{d}N_{\text{pair}}}{\text{d}\dphi} = 
 \frac{1}{\deta_{b}-\deta_{a}} \int_{\deta_{a}}^{\deta_{b}} 
 \frac{1}{N_{\text{trig}}} \frac{\text{d}^{2}N_{\text{pair}}}{\text{d}\deta \text{d}\dphi} \text{d}\deta.
\end{equation}
The short-range correlations, \eg of the jet-peak, are excluded by averaging the two-dimensional yield over the interval from $\deta_{a}=2.0$ to $\deta_{b}=2.9$.
Since random particle combinations produce a flat pedestal in the yield, the correlation structures of interest are extracted by using the zero-yield-at-minimum (ZYAM) method~\cite{Adler:2002tq,PhysRevC.72.011902}.
By fitting a second-order polynomial to $Y(\dphi)$ in the range $0.1<\dphi<2.0$, the offset is estimated as the minimum of the polynomial.
This value, further denoted as $C_{\scriptscriptstyle{\text{ZYAM}}}$, is subtracted from $Y(\dphi)$ to shift its minimum to be at zero yield. 
The uncertainties on $C_{\scriptscriptstyle{\text{ZYAM}}}$ due to the limited sample size and the fit range are below $0.002$ for all individual measurements.

\begin{figure}[!p]
  \begin{center}
    \includegraphics[width=0.75\linewidth]{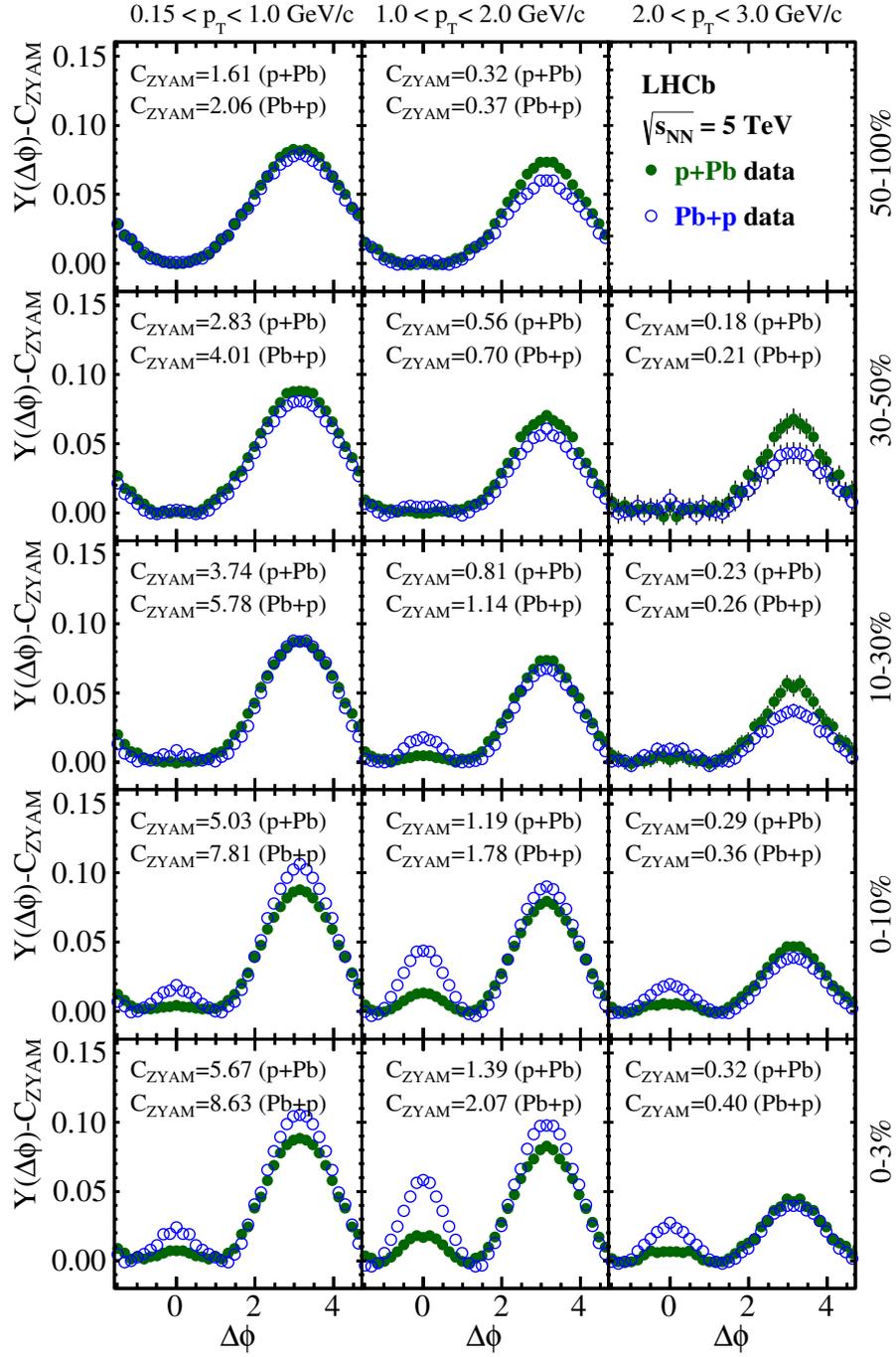}    
  \end{center}
  \caption{\small 
  One-dimensional correlation yield as a function of $\dphi$ obtained from the ZYAM-method by averaging over $2.0<\deta<2.9$. 
  The subtracted yields are presented for $\sqrtnn=5\mathrm{\,Te\kern -0.1em V} \xspace$ proton-lead collisions recorded in \ppb (full green circles) and \pbp (open blue circles) configurations. 
  The ZYAM constant is given in each panel.
  Event classes are compared for low to very-high activities from top to bottom, and different intervals of increasing $\pt$ from left to right.
  Only statistical uncertainties are shown. 
  Error bars are often smaller than the markers.
  }
  \label{fig:cor1D_pA}
\end{figure}

The subtracted one-dimensional yields for the \ppb (full circles) and \pbp (open circles) data samples are shown in Fig.~\ref{fig:cor1D_pA} for all activity classes and \pt intervals. 
The correlation increases with event activity, but decreases towards higher \pt where fewer particles are found. 
Since more particles are emitted into the acceptance of the detector in the \pbp compared to the \ppb configuration, a larger offset is observed, as indicated by the ZYAM constants.
All distributions in Fig.~\ref{fig:cor1D_pA} show a maximum at $\dphi=\pi$, marking the centre of the away-side ridge, which balances the momentum of the near-side (the jet peak is excluded in this representation).
The lower activity classes, $50-100\%$ and $30-50\%$, do not have a corresponding maximum at $\dphi=0$.
The $30-50\%$ event class of the \pbp sample shows a first change in shape of the distribution at $\dphi=0$. 
The picture changes when probing the intermediate activity class $10-30\%$.
In all \pt intervals of the \pbp sample the emergence of the near-side ridge with a second maximum at $\dphi=0$ is clearly visible.
In the \ppb sample the event activity is still not high enough to form a clear near-side structure.
In the high-activity classes, $0-10\%$ and $0-3\%$, the near-side ridge is strongly pronounced in the \pbp sample in all \pt intervals. 
In the \ppb sample the near-side structure is less distinct; however the $1<\pt<2\gevc$ interval shows a clear near-side ridge.

A qualitatively similar behaviour is seen in the forward-central correlations studied by the \alice experiment~\cite{Adam:2015bka}, with a forward muon trigger and central associated particles. 
Here also a clear ridge effect is observed, which grows with increasing event activity, and indications are seen that it is more pronounced in the hemisphere of the Pb nucleus.

Comparison of the ZYAM-subtracted yields shows that the away-side ridge is always more prominent than the near-side ridge. 
The ridge on the away-side is only weakly dependent on \pt, while the near-side ridge appears most pronounced in the bin $1<\pt<2\gevc$. 
Comparing \ppb and \pbp, one finds that especially for high event activities the near-side ridge is more pronounced in the Pb hemisphere.

Study of the one-dimensional yields within a \pt interval for different activity classes shows that the away side remains approximately unchanged, while the near side starts to form the additional ridge when a certain event activity is reached.
This turn-on, however, appears to be at different activities in the \ppb and \pbp configurations.

The same qualitative observations in the various analysis bins, including the emergence of the near-side ridge, are found when using the track-based approach for the definition of the activity classes as a systematic check.
The total correlation yield varies by only a few percent, and the maximum variation does not exceed $10\%$ in the low-\pt range.
The emergence of the near-side ridge in the ZYAM-subtracted yield is unaffected by the change of the event activity definition.

Further systematic effects related to the event selection are evaluated by including events with multiple reconstructed primary vertices. 
The change of the final correlation yield is negligible. 
As another cross-check, data recorded in magnet up and down polarities are analysed separately. 
The results are in good agreement with each other.

\begin{figure}[tb]
\centering
  \includegraphics[width=1.\linewidth]{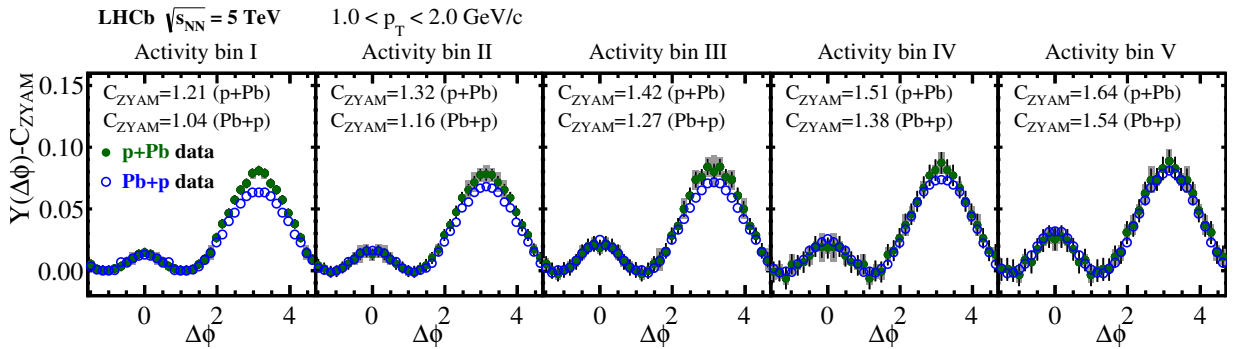}\\
  \caption{
  \small
  One-dimensional correlation yield as a function of $\dphi$ obtained from the ZYAM-method by averaging the two-dimensional distribution over $2.0<\deta<2.9$. 
  The results for the \ppb and \pbp samples are compared in five event classes which probe identical activities in the range $2.0<\eta<4.9$. 
  The measured hit-multiplicities of the \ppb sample are scaled to agree with the hit-multiplicities of the \pbp sample. 
  The uncertainty band represents the systematic limitation of the scaling procedure.
  The error bars represent the statistical uncertainty.
  }
  \label{fig:cor1D_abs}
\end{figure}

To investigate the activity dependence of the long-range correlations in the \ppb and \pbp samples in more detail, common bins in absolute activity for both samples are studied.
For this purpose, events of both samples are probed in which a similar number of charged particles are emitted into the forward direction.
Events of both samples are grouped into five narrow activity bins, as defined in Table~\ref{tab:AbsActivityBins}.
Figure~\ref{fig:cor1D_abs} compares the ZYAM-subtracted two-particle correlation yields in the range $1<\pt<2\gevc$, in which the near-side ridge is most pronounced.
The uncertainty bands represent the systematic uncertainty on the scaling factor, which translates the activity of the \ppb configuration to that of the \pbp configuration.
For \ppb and \pbp events of the same activity in the forward region, the observed long-range correlations become compatible within the uncertainties, except for bin I in which the away-side yield in \ppb is still slightly more pronounced.
The near-side correlation in the beam (\textit{p}) and target (Pb) fragmentation hemispheres shows a consistent increase with increasing event activity.

\section{Summary and conclusions}
\label{sec:Conclusion}

Two-particle angular correlations between prompt charged particles produced in \pa collisions at $\sqrtnn=5\tev$ have been measured for the first time in the forward region, using the \lhcb detector.
The angular correlations are studied in the laboratory frame in the pseudorapidity range $2.0<\eta<4.9$ over the full range of azimuthal angles, probing particle pairs in different common \pt intervals.
With the asymmetric detector layout, the analysis is performed separately for the \ppb and \pbp beam configurations, which probe rapidities in the nucleon-nucleon centre-of-mass frame of $1.5<y<4.4$ and $-5.4<y<-2.5$, respectively. 
The strength of the near-side ridge observed in the backward (\pbp configuration) region appears to be of similar size to that found in the forward (\ppb configuration) region.
The relative shift of about one unit in nucleon-nucleon centre-of-mass rapidity between the two configurations produces no sizeable effect on the near-side ridge within the accuracy of the measurement.
For events with high event activity a long-range correlation on the near side (the ridge) is observed in both configurations.
While the correlation structure on the away side shrinks with increasing \pt, the near-side ridge is most pronounced in the range $1<\pt<2\gevc$.
The observation of the ridge in the forward region extends previous \lhc measurements, which show similar qualitative features.
Furthermore, the correlation dependence on the event activity is investigated for relative and absolute activity ranges.
The correlation structures on the near side and on the away side both grow stronger with increasing event activity.
For identical absolute activity ranges in the \ppb and \pbp configurations the observed long-range correlations are compatible with each other.

\section*{Acknowledgements}
 
\noindent We express our gratitude to our colleagues in the CERN
accelerator departments for the excellent performance of the LHC. We
thank the technical and administrative staff at the LHCb
institutes. We acknowledge support from CERN and from the national
agencies: CAPES, CNPq, FAPERJ and FINEP (Brazil); NSFC (China);
CNRS/IN2P3 (France); BMBF, DFG and MPG (Germany); INFN (Italy); 
FOM and NWO (The Netherlands); MNiSW and NCN (Poland); MEN/IFA (Romania); 
MinES and FANO (Russia); MinECo (Spain); SNSF and SER (Switzerland); 
NASU (Ukraine); STFC (United Kingdom); NSF (USA).
We acknowledge the computing resources that are provided by CERN, IN2P3 (France), KIT and DESY (Germany), INFN (Italy), SURF (The Netherlands), PIC (Spain), GridPP (United Kingdom), RRCKI (Russia), CSCS (Switzerland), IFIN-HH (Romania), CBPF (Brazil), PL-GRID (Poland) and OSC (USA). We are indebted to the communities behind the multiple open 
source software packages on which we depend. We are also thankful for the 
computing resources and the access to software R\&D tools provided by Yandex LLC (Russia).
Individual groups or members have received support from AvH Foundation (Germany),
EPLANET, Marie Sk\l{}odowska-Curie Actions and ERC (European Union), 
Conseil G\'{e}n\'{e}ral de Haute-Savoie, Labex ENIGMASS and OCEVU, 
R\'{e}gion Auvergne (France), RFBR (Russia), GVA, XuntaGal and GENCAT (Spain), The Royal Society 
and Royal Commission for the Exhibition of 1851 (United Kingdom).

\addcontentsline{toc}{section}{References}
\setboolean{inbibliography}{true}
\bibliographystyle{LHCb}
\bibliography{main_PLB3,LHCb-PAPER,LHCb-CONF,LHCb-DP,LHCb-TDR}

\newpage

\centerline{\large\bf LHCb collaboration}
\begin{flushleft}
\small
R.~Aaij$^{39}$, 
C.~Abell\'{a}n~Beteta$^{41}$, 
B.~Adeva$^{38}$, 
M.~Adinolfi$^{47}$, 
A.~Affolder$^{53}$, 
Z.~Ajaltouni$^{5}$, 
S.~Akar$^{6}$, 
J.~Albrecht$^{10}$, 
F.~Alessio$^{39}$, 
M.~Alexander$^{52}$
% ...

{\footnotesize \it
% ...
$ ^{5}$Clermont Universit\'{e}, Universit\'{e} Blaise Pascal, CNRS/IN2P3, LPC, Clermont-Ferrand, France\\
$ ^{6}$CPPM, Aix-Marseille Universit\'{e}, CNRS/IN2P3, Marseille, France\\
% ...
$ ^{10}$Fakult\"{a}t Physik, Technische Universit\"{a}t Dortmund, Dortmund, Germany\\
% ...
$ ^{38}$Universidad de Santiago de Compostela, Santiago de Compostela, Spain\\
$ ^{39}$European Organization for Nuclear Research (CERN), Geneva, Switzerland\\
% ...
$ ^{41}$Physik-Institut, Universit\"{a}t Z\"{u}rich, Z\"{u}rich, Switzerland\\
% ...
$ ^{47}$H.H. Wills Physics Laboratory, University of Bristol, Bristol, United Kingdom\\
% ...
$ ^{52}$School of Physics and Astronomy, University of Glasgow, Glasgow, United Kingdom\\
$ ^{53}$Oliver Lodge Laboratory, University of Liverpool, Liverpool, United Kingdom\\
% ...
}
\end{flushleft}

\end{document}